\documentclass[aps,prb,twocolumn,floatfix,superscriptaddress,amsmath,amssymb]{revtex4}
\usepackage[varg]{txfonts}
\usepackage[final]{graphicx}
\usepackage{amsmath,amssymb,bbold,bm}
\newcommand\vex[1]{\mathbf{#1}}

\def\id{\mathbb{1}}

\def\vex#1{\mathbf{#1}}
\def\ii{\mathrm{i}}
\renewcommand{\Re}{\mathrm{Re}}

\def\slash#1{\setbox0=\hbox{$#1$}			
   \dimen0=\wd0                                			
   \setbox1=\hbox{/} \dimen1=\wd1  	 		
   \ifdim\dimen0>\dimen1                               	
      \rlap{\hbox to \dimen0{\hfil / \hfil}} 	  	
      #1                                      				
   \else                                        				
      \rlap{\hbox to \dimen1{\hfil$#1$\hfil}}   		
      \hbox{/} 	                              			
   \fi}   
   
\begin{document}
\title{Fractionalization in a square-lattice model with time-reversal symmetry}
\author{B. Seradjeh}
\affiliation{Department of Physics and Astronomy, University of British Columbia, Vancouver, BC, Canada V6T 1Z1}
\author{C. Weeks}
\affiliation{Department of Physics and Astronomy, University of British 
Columbia, Vancouver, BC, Canada V6T 1Z1}
\author{M. Franz}
\affiliation{Department of Physics and Astronomy, University of British 
Columbia, Vancouver, BC, Canada V6T 1Z1}
\affiliation{Kavli Institute for Theoretical Physics, University of 
California, Santa Barbara, CA 93106}
\begin{abstract}
We propose a two-dimensional time-reversal invariant system of essentially
non-interacting electrons on a square lattice that exhibits configurations with 
fractional charges $\pm e/2$. These are vortex-like topological defects 
in the dimerization order parameter describing spatial modulation in the electron 
hopping amplitudes. Charge fractionalization is established by a simple 
counting argument, analytical calculation within the effective low-energy theory, 
and by an exact numerical diagonalization of the lattice Hamiltonian. We comment 
on the exchange statistics of fractional charges and possible realizations of 
the system.
\end{abstract}
\maketitle

\emph{Introduction.}---It is now well-known that fractional quantum numbers can 
arise as the collective excitations of a many-body system. The canonical example 
of such fractionalization is a 2D electron gas (2DEG) placed in a transverse
 magnetic field in the fractional quantum Hall regime. At odd inverse filling 
factors, $\nu^{-1}>1$, the many-body ground state is described by a 
strongly-correlated Laughlin wave function and the time-reversal symmetry is 
broken. The excitations carry the fractional charge $\nu e$ 
(Ref.~\onlinecite{Lau83b}) and exhibit fractional (Abelian) 
statistics.~\cite{AroSchWil84a} The search for other systems that exhibit 
fractionalization is ongoing. Two important questions in this search are whether 
strong correlations or a broken time-reversal symmetry is necessary for 
fractionalization to happen.

Recently the answer to both questions is argued to be negative. A group including the present authors~\cite{WeeRosSer07a} proposed a system with fractionally 
charged, anyonic excitations that can be described by a weakly-interacting wave 
function found by filling a set of single-particle states. Moreover, Hou, Chamon 
and Mudry~\cite{HouChaMud07a} have argued that a vortex in the Kekul\'e
 modulations of the hopping amplitudes on a honeycomb lattice, like that of 
graphene, binds a fractional charge $e/2$ without breaking the time resversal 
symmetry.

In this paper we propose a system on a square lattice with time-reversal symmetry that exhibits fractionalization. The system consists of a square lattice threaded by one half of a magnetic flux quantum $\Phi_0=hc/e$ per plaquette on which electrons can hop to nearest-neighbor sites with no interaction. Time-reversal symmetry is preserved in this lattice model because electrons cannot detect the sign of the flux with magnitude $\Phi_0/2$. In addition, we assume a dimerized modulation of hopping amplitudes as depicted in Fig. 1(a). Such modulations can arise as a Peierls distortion of a uniform ion lattice above a critical value of the electron-phonon coupling or, as discussed 
in Ref.~\onlinecite{HouChaMud07a} in the context of graphene, as an interaction-driven instability. We show that a vortex in the complex scalar order parameter describing this dimerization pattern generates a zero-energy bound state in the spectrum of electrons and carries a fractional charge. In addition to similar arguments to those of Ref.~\onlinecite{HouChaMud07a}, we present a simple electron counting argument\cite{GolWil81a,SuSchHee79a} as well as numerical evidence demonstrating this effect. Importantly, we find that the fractionalization survives essentially intact beyond the low-energy theory of Ref.~\onlinecite{HouChaMud07a} in the presence of the lattice. We also show that the same pattern of fractionalization occurs for the $Z_4$ and U(1) vortices and clarify the issue of their confinement. 
\begin{figure}[tb]
\begin{center}
\includegraphics[width=0.45\textwidth]{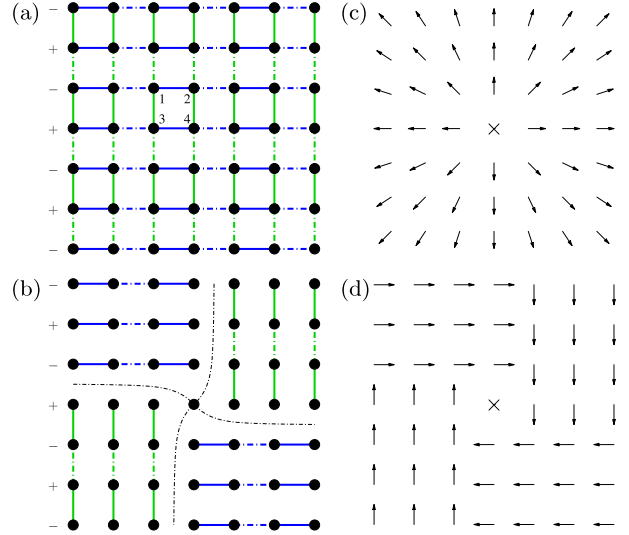}
\caption{(Color online) The model: (a) Square lattice with $\frac12\Phi_0$ 
magnetic flux per plaquette and dimerized hopping amplitudes. The $\pm$ on the 
left for each row show the choice of gauge for the Peierls phase factors. The solid (dashed) bonds indicate an increased (decreased) hopping amplitude in the $\hat x$ (blue) and $\hat y$ (green) directions as explained in the text. The four sites of the 
unit cell are marked. (b) The $Z_4$ vortex. The dashed lines indicate the domain 
walls sharing the center of the vortex. The phase of the local dimer order parameter 
$f_i$ (see text) around (c) the U(1) and (d) the $Z_4$ vortex. The $\times$ 
shows the center of each vortex where $f=0$.}
\label{fig:SqL}
\end{center}
\end{figure}
We then give a brief discussion of the energetics of the Peierls distortion and touch upon possible experimental realizations of the model in artificially engineered semiconductor heterostructures and optical lattices. 

\emph{The model.}---We consider a square lattice with a tight-binding Hamiltonian
\begin{equation}\label{eq:tbH}
\mathcal{H}_e = -\sum_{\left<ij\right>}\left(t_{ij}e^{\ii\theta_{ij}}c^\dag_jc_i+
\mathrm{h.c.}\right)+\sum_i \epsilon_i c^\dag_ic_i,
\end{equation}
for electrons annihilated at site ${\vex r}_i$ by $c_i$. We assume the spins of all
electrons are polarized along the field. Here, $t_{ij}$ are hopping amplitudes 
between nearest-neighbor sites and $\epsilon_i$ is the on-site potential. The magnetic field is included through the Peierls phase factors,
$ 
\theta_{ij} = (2\pi/\Phi_0)\int_{\vex r_i}^{\vex r_j} \vex A\cdot d\vex l.
$ 
Each plaquette is threaded by flux $\Phi_0/2$. We choose 
to work in the Landau gauge $\vex A=(\Phi_0/2)(-y,0)$ where we have set the 
lattice spacing to unity. 
The hopping amplitudes in our
model form a dimerized pattern as shown in Fig.~\ref{fig:SqL}(a) with 
alternating bonds $t_{i,i+\hat{x}}=t(1\pm m_x)$ and $t_{i,i+\hat{y}}=t(1\pm m_y)$. 
In the following we assume uniform $\epsilon_i=\epsilon$. 

We can now arrange the four sites of the unit cell into a spinor field
\begin{equation}
\psi_{\vex k} = e^{\frac{\pi}4\gamma_2}\left( c_{1{\vex k}},c_{2{\vex k}},c_{3{\vex k}},c_{4{\vex k}} \right)^T
\end{equation}
to write the Hamiltonian as
\begin{equation}
\mathcal{H}_e = \sum_{\vex k\in \mathrm{BZ}}\psi^\dag_{\vex k}\left(\epsilon-2tH_{\vex k}\right)\psi_{\vex k}
\end{equation}
in the reduced Brillouin zone $\mathrm{BZ}=\{\vex k \big| |k_x|, |k_y| \leq \frac12\pi\}$ with
\[
H_{\vex k} = \gamma_0\left(\gamma_1\cos k_x + \gamma_2 \cos k_y + m_x \sin k_x+\ii m_y \gamma_5 \sin k_y\right).
\]
We are using the standard Weyl representation of Dirac matrices:
$\gamma_0=\sigma_1\otimes\id, \vec\gamma = \ii\sigma_2\otimes\vec\sigma$ and 
$\gamma_5=-\ii\gamma_0\gamma_1\gamma_2\gamma_3=\sigma_3\otimes\id$.  The spectrum of $H_{\vex k}$ is given by
\begin{equation}\label{eq:Ek}
E_{\vex k}=\pm\sqrt{\cos^2k_x+\cos^2k_y+m_x^2\sin^2k_x+m_y^2\sin^2k_y},
\end{equation}
which is doubly degenerate and symmetric: if $\psi_E$ is an eigenstate with 
energy $E$, i.e. $H_{\vex k}\psi_E=E\psi_E$, then $\gamma_0\gamma_3\psi_E$ is an 
eigenstate with energy $-E$ since $\{H_{\vex k},\gamma_0\gamma_3\}=0$.

\emph{Low-energy theory.}---The undistorted system with $\vex m\equiv(m_x,m_y)=0$ 
has 
Dirac nodes at $k_x=k_y=\pm\pi/2$ where the two energy bands meet. There is one 
node per reduced Brillouin zone. At nonzero constant dimerization $\vex m$ a gap 
$m=(m^2_x+m^2_y)^{1/2}$ opens up in the spectrum. We will focus our attention on 
the low-energy excitations of the system in the situation when the dimerization
vector ${\vex m}$ is slowly varying in space. Using a polar representation for the
hopping anisotropy $m_x+\ii m_y=me^{\ii\chi}$ and by linearizing $\mathcal{H}_e$ 
around the nodes in the continuum limit we find the low-energy Hamiltonian
\begin{equation}\label{eq:cHx}
H_{\vex x}=\gamma_0\left(\ii\gamma_1 \partial_x+\ii\gamma_2 
\partial_y+me^{\ii\chi\gamma_5}\right).
\end{equation}

An interesting situation arises when a U(1) vortex is present in the dimerization
vector ${\vex m}$. Such a vortex is characterized by a $2n\pi$ winding in the phase 
$\chi$, with $n$ an integer, along a contour enclosing the vortex center. 
The continuum Dirac 
Hamiltonian~(\ref{eq:cHx}) in this vortex background field is the same as that 
studied in Refs.~\onlinecite{HouChaMud07a} and~\onlinecite{JacPi07a}. It admits 
$|n|$ degenerate 
zero-energy bound states.~\cite{JacRos81a} This result has its origin in an
index theorem for the Dirac Hamiltonian in topologically nontrivial background fields.~\cite{Wei81a} 

For a singly quantized vortex ($|n|=1$) there is a single 
zero mode with support on one of the sublattices depending on the sense of the 
vortex. A standard calculation\cite{HouChaMud07a,JacReb76a} can be used to show 
that when the lattice is half-filled the charge carried by such a vortex is {\em fractionalized}, namely $\pm e/2$, depending on whether or not the zero mode is occupied by an electron. 

In our square lattice model one can also give a simple intuitive counting argument for the charge fractionalization that generalizes the  construction deployed in the case 
of polyacetylene by Goldstone and Wilczek.~\cite{GolWil81a} Consider a $Z_4$ vortex,
depicted in Fig.\ 1(b,d), in the limit of strong dimerization, $m\simeq 1$. In this limit each thick bond in Fig.\ 1(b) can 
be thought of as representing one electron resonating between the two lattice 
sites. The topological structure of this Z$_4$ vortex guarantees that in
its core there is necessarily a site with no partner.~\cite{LevSen04a} This
site can be either occupied or empty, thus exhibiting $e/2$ surplus or deficit
of charge compared to the background, which has $e/2$ per site on average.
The $Z_4$ vortex can be adiabatically deformed into the U(1) vortex of Fig.\ 1(b)
and to the weakly dimerized phase at $m\ll 1$. 
This indicates that the charge fractionalization persists in that limit, as 
explicitly verified by the analytical calculation\cite{HouChaMud07a,JacReb76a} 
and by the numerical results reported below. We note that this argument works in 
our model only in the presence of the magnetic flux, which is needed for 
the existence of the Dirac nodes in the spectrum.
At zero flux there is a Fermi surface for any small $m$ and the 
unpaired center of a vortex does not create a bound state.

\emph{Numerics.}---We have performed exact diagonalization of the tight-binding 
Hamiltonian~(\ref{eq:tbH}) with different vortex configurations on lattices of 
size up to $52\times 52$. Open boundary conditions were assumed in all cases. We 
consider two realizations of the vortex, a $Z_4$ vortex and a lattice version of 
the continuum U(1) vortex studied in the low-energy theory. To quantify the vortex
structure we use a local dimer order parameter~\cite{ReaSac90a}  defined at 
site ${\vex r}_i=(x_i,y_i)$ as
\begin{equation}
f_i = \sum_{\hat\mu}\eta_{i,i+\hat\mu} t_{i,i+\hat\mu}.
\end{equation}
Here the $\hat\mu$ are the four nearest-neighbor unit vectors, 
$\eta_{i,i+\hat x}=(-1)^{x_i}$, $\eta_{i,i+\hat y}=\ii(-1)^{y_i}$ and 
$\eta_{i,i-\hat\mu}=\eta_{i-\hat\mu,i}$. (See also Fig.~2 in Ref.~\onlinecite{ReaSac90a}.)

The $Z_4$ vortex occurs if the energetics of the problem allow only dimerizations
along the high-symmetry directions of the crystal, e.g.\ ${\vex m}=(\pm m,0)$ and 
$(0,\pm m)$. A vortex is then obtained
by putting together four possible domains as illustrated in Fig.~\ref{fig:SqL}(b),
separated by domain walls emanating from 
the vortex center. The phase of the order parameter for the $Z_4$ vortex
is depicted in Fig.~\ref{fig:SqL}(d).

The U(1) vortex occurs if the energy of the state characterized by dimerization 
vector of the form ${\vex m}=m(\cos\chi,\sin\chi)$ is approximately independent 
of $\chi$. We construct the U(1) vortex on the lattice by discretizing the 
continuum vortex field $\varphi(z)=z/|z|$ where $z=x+\ii y$ is the complex 
coordinate of the point $(x,y)$. Explicitly,
\begin{equation}\label{eq:tU1}
t_{i,i+\hat\mu}=t_0\big(1+m\Re\left[\eta_{i,i+\hat\mu}^*\varphi
\left({\vex r}_i+\mbox{$\frac12$}\hat\mu\right)\right]\big),
\end{equation}
where the asterisk denotes the complex conjugation. Fig.~\ref{fig:SqL}(b) shows 
the phase of $f_i$ clearly illustrating the U(1) form. 
The U(1) structure can be extended to study the case for multiple vortices
 and antivortices. For vortices at $z_1, z_2, \dots$ and antivortices at 
$z'_1, z'_2, \dots$ this is achieved by taking a product of vortex fields 
$\varphi(z)=\prod_a(z-z_a)/|z-z_a|\prod_b(z-z'_b)^*/|z-z'_b|$ in 
Eq.~(\ref{eq:tU1}). 

\begin{figure}[t]
\begin{center}
\includegraphics[width=0.45\textwidth]{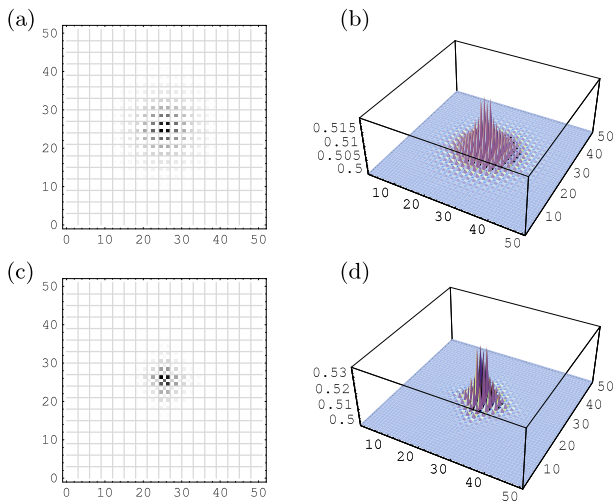}
\caption{(Color online) The charge density at half-filling with zero mode 
filled  in the U(1) vortex with $m=0.15$ (a,b).  The same plots for the $Z_4$ 
vortex and $m=0.30$ (c,d). The lattice size $51\times 51$ and the gridline spacing is 3. The top-view density 
plots on the left are included to show the symmetry of the solution for each 
vortex and the support on a single sublattice.}
\label{fig:1v}
\end{center}
\end{figure}
\begin{figure}[t]
\begin{center}
\includegraphics[width=0.45\textwidth]{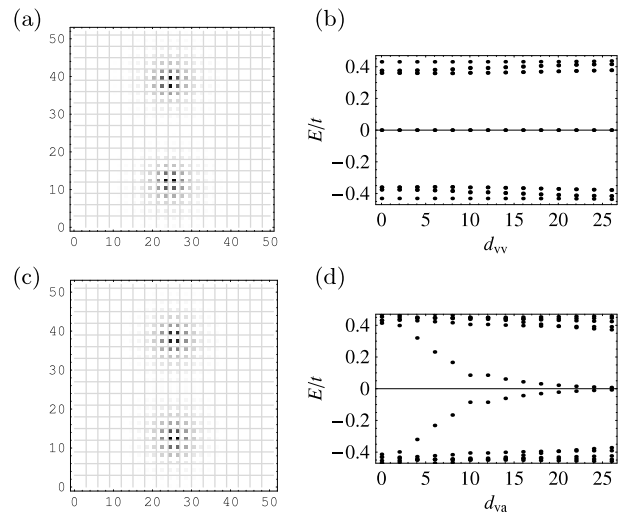}
\caption{The charge density at half-filling with both midgap states filled (a,c) and the 
low-energy (including midgap) eigenvalues (b,d) of a system with two U(1) vortices (a,b)
and with a vortex and an antivortex (c,d). The gap $m=0.2$, the lattice size is 
$52\times51$ and the gridline spacing is 3.}
\label{fig:2v}
\end{center}
\end{figure}

A typical sample of our results for a system with one and two vortices is shown in 
Figs.~\ref{fig:1v} and~\ref{fig:2v}, respectively. The structure of the vortex is 
reflected in the symmetries of the zero mode and the corresponding charge density 
at half-filling. For a single vortex, Fig.~\ref{fig:1v}, the charge density at 
half-filling with the zero mode filled 
shows an excess charge localized at the vortex above the background value. For 
both the U(1) and $Z_4$ vortex this 
excess charge integrates to $0.5e$ to within machine accuracy confirming the 
existence of the fractional charge. Remarkably, the charge density of the half-filled system with the $Z_4$ vortex is essentially the same as that with the U(1) vortex. Especially, no charge is bound to the domain walls far away from the center.

The system with two vortices has two zero-energy states with support on the same 
sublattice, Fig.~\ref{fig:2v}(a). It is important to note that, as illustrated in
Fig.~\ref{fig:2v}(b), these remain {\em exact zero modes} even when the 
intervortex separation $d_{\mathrm{vv}}$ becomes short compared to the size of 
the bound state wavefunction $\xi\sim m^{-1}$. The zero modes are topologically
protected in accordance with the index theorem of Ref.~\onlinecite{Wei81a} which 
states that the number of exact zero modes in the system is bounded from below by 
the total vorticity of the order parameter $f$. This can be understood on a more
intuitive level by noting that since the two bound states reside on the same
sublattice the matrix element of the Hamiltonian (\ref{eq:tbH}) with $\epsilon=0$
between the two states vanishes identically and hence leads to no splitting in
energies. Alternatively, one can consider a process of merging
$n$ single vortices into one vortex with winding $n$. According to the Ref.~\onlinecite{JacRos81a} the latter supports $|n|$ zero modes. Thus, the zero modes
associated with $n$ individual widely separated vortices transform smoothly
into $n$ zero modes of the $n$-fold vortex.
 
The vortex-antivortex configuration, by contrast, does not have exact zero modes.
 Now the two bound states reside on different sublattices, Fig.~\ref{fig:2v}(c), 
so there is a finite tunneling amplitude between them. This yields two midgap 
states with energies $\pm E_{\mathrm{va}}$ that depend exponentially on the 
vortex-antivortex separation $d_{\mathrm{va}}$ and merge with the continuum states
as $d_{\mathrm{va}}\to0$, Fig.~\ref{fig:2v}(c). Again, these results are consistent 
with the index theorem\cite{Wei81a} since the total vorticity of a 
vortex-antivortex pair vanishes.

\emph{Peierls distortion.}---In a crystalline system the dimerizations postulated 
in our model may occur through a Peierls distortion of the ion lattice in the presence of the magnetic flux at or near 
half-filling. Such a distortion is driven by the electron kinetic energy gain upon
the opening of a gap. We briefly outline this mechanism. At half filling, the energy gain per site is 
$ 
\delta\mathcal{E}_e=-\frac{4t}{N}\sum_{\vex k}[E_{\vex k}(\vex m)-E_{\vex k}(0)], 
$ 
where $N$ is the number of sites and $E_{\vex k}(\vex m)$ is given by the positive
branch of Eq. (\ref{eq:Ek}).
This gain is to be balanced against the loss of elastic energy that results
from the ion  displacements $\vex u_i$, 
$ 
\delta\mathcal{E}_{\mathrm{latt}}=\frac{K}{2N}\sum_{\langle ij\rangle}(\vex u_i-\vex u_j)^2.
$ 
We assume henceforth that 
the change in the hopping element is related to the displacement  as
$m_{i,i+\hat\mu}=g\hat\mu\cdot(\vex u_i-\vex u_{i+\hat\mu})$ with $g$ the dimensionless
 electron-phonon coupling.\cite{TanHir88a} For the dimerization pattern
 characterized by a 
constant, spatially uniform vector ${\vex m}$, we have 
$\delta\mathcal{E}_{\mathrm{latt}}/4t=\frac12\kappa m^2$ with $\kappa=K/4tg^2$. 

The Peierls distortion occurs when $\delta\mathcal{E}=\delta\mathcal{E}_e+
\delta\mathcal{E}_{\mathrm{latt}}$ becomes negative. Unlike the one-dimensional case, 
where the Peierls instability occurs at infinitesimal $g$, in two dimensions $g$ must exceed a critical value. In terms of $\kappa$ this happens 
for $\kappa<\kappa_c= N^{-1}\sum_{\vex k}\sin^2k_x/E_{\vex k}(0)\simeq 0.806$. Near 
the transition we can expand
\begin{equation}\label{eq:expandE}
\frac{\delta\mathcal{E}}{4t}={\frac12}(\kappa-\kappa_c)m^2+\frac2{3\pi}|m|^3
+{\cal O}(m^4).
\end{equation}
We observe that the leading terms in $\delta\mathcal{E}$ are isotropic in $\vex m$. 
The anisotropy first enters at the fourth order via a term $\sim(m_x^2-m_y^2)^2$. This 
indicates that for small $m$ a vortex in this model will closely resemble the U(1) 
vortex in Fig.\ 1(c), with a small four-fold anisotropy. For $m\simeq 1$ 
the anisotropy becomes important and the pattern will be closer to a $Z_4$ 
vortex, Fig.\ 1(d). 

These findings are important since the $Z_4$ vortices are generically confined by a linear potential due to the domain walls while the U(1) vortices are not. For the fractional charges to be observable we require that the $Z_4$ confinement length scale be much larger than $\xi$. It is also possible that the U(1) symmetry emerges universally at the dimerization transition. We note that in this case the small $Z_4$ anisotropy of the present model is an irrelevant perturbation while the $Z_3$ in graphene~\cite{HouChaMud07a} is relevant.

\emph{Experimental realization.}---In a natural solid 
with lattice constant $a_0\approx2$\AA \ the magnetic field required to achieve 
the $\Phi_0/2$ flux per plaquette is of the order $10^4$\ T  and thus
much too high to reach in a laboratory. A better chance for observing 
the phenomena discussed here is to artificially engineer the hosting system
such that the electrons move in a lattice with much larger $a_0$. 

Indeed several
schemes have been proposed and implemented recently to achieve this effect in
semiconductor heterostructures (such as GaAs/AlGaAs) combined with a type-II
superconducting layer\cite{WeeRosSer07a,BerRapJan05a} or a top gate realized
by the diblock copolymer nanolithography technique.\cite{melinte1} 
A second possibility is offered by fermionic cold atoms in an optical lattice 
with artificial magnetic flux.\cite{JakZol03a} Once the tight binding model
with the magnetic flux is realized by any of the above techniques
there of course remains the challenge of creating a vortex and 
detecting the fractional charge. 

\emph{Conclusion.}---We showed through general arguments and a detailed numerical 
study that the system proposed here supports configurations with  fractional charge while preserving time-reversal symmetry. The number of zero modes in vortex configurations leading to this fractionalization is topologically protected. This should be contrasted with similar zero modes in vortices of a chiral $p$-wave superconductor where such a generic topological protection seems not to exist.~\cite{GurRad07a} 
An interesting question is what exchange statistics these fractional particles obey. This is significant theoretically as well as for potential applications in quantum information processing. A natural expectation, based on the analogy with the fractional quantum Hall states,\cite{Lau83b,AroSchWil84a} is that they are anyons. This is also suggested by arguments using gauge invariance on a torus.~\cite{WuHatKoh91a}  Since the system is time-reversal invariant one expects the effective theory for the topological excitations to be a ``doubled'' Chern-Simons gauge theory of the type discussed in Ref.~\onlinecite{FreNaySht04a}. Indeed, our preliminary investigation~\cite{SerFra07x} suggests that such a gauge structure will emerge upon integrating out the Fermi fields in the Dirac Hamiltonian (\ref{eq:cHx}) in the presence of topologically nontrivial background field $\chi$.

\emph{Acknowledgment.}---The authors have benefited from discussions and 
correspondence with L. Balents, M.P.A. Fisher, V. Gurarie, R. Jackiw, K. Madison, 
G. Rosenberg, J. Pachos and A. Vishwanath. This work has
been supported by the CIAR, NSERC, NSF Grant No. PHY05-51164 (KITP) 
and the Killam Foundation.

\end{document}